# Equatorial trench at the magnetopause under saturation

A. Dmitriev[1,2] and A. Suvorova[3,2]

[1]*Institute of Space Science, National Central University, Chung-Li, Taiwan*
[2]*Institute of Nuclear Physics, Moscow State University, Moscow, Russia*
[3]*Center for Space and Remote Sensing Research, National Central University, Chung-Li, Taiwan*

Short title: MAGNETOPAUSE TRENCH

**Abstract** Magnetic data from GOES geosynchronous satellites were applied for statistical study of the low-latitude dayside magnetopause under a strong interplanetary magnetic field of southward orientation when the reconnection at the magnetopause was saturated. From minimum variance analysis, we determined the magnetopause orientation and compared it with predictions of a reference model. The magnetopause shape was found to be substantially distorted by a duskward shifting such that the nose region appeared in the postnoon sector. At equatorial latitudes, the shape of magnetopause was characterized by a prominent bluntness and by a trench formed in the postnoon sector. The origin of distortions was regarded in the context of the storm-time magnetospheric currents and the large-scale quasi-state reconnection at the dayside magnetopause.

## 1. Introduction

It is widely accepted that the subsolar magnetopause moves earthward due to the reconnection of geomagnetic field with a southward component (negative $Bz$) of interplanetary magnetic field (IMF). Stronger southward IMF results in more intense reconnection and, thus, smaller distance to the magnetopause. However, when the magnitude of negative IMF $Bz$ exceeds a certain threshold of about 20 nT, the magnetopause distance does not decrease any more, i.e. the southward IMF reconnection is saturated [e.g. *Yang et al.*, 2003; *Suvorova et al.*, 2005; *Dmitriev et al.*, 2011]. Statistical studies indicate that the IMF $Bz$ threshold for saturation changes from about -20 nT to about -10 nT with increasing geomagnetic activity [*Dmitriev et al.*, 2011]. The effect of reconnection saturation is still poorly investigated because of limited amount of experimental data.

The saturation effect is studied using global MHD simulations. *Siscoe et al.* [2004] show that at the stagnation point (subsolar magnetopause), the region 1 field-aligned currents (FAC) cause depletion of the geomagnetic field as well as substantial changing of the shape of magnetopause such that a dimple develops at the stagnation point and the magnetopause becomes blunt and the bow shock recedes. These effects lead to decrease and limitation of the reconnection at the dayside magnetopause. Analyzing geomagnetic field observations during the strong southward IMF, *Ober et al.* [2006] have found an indirect support of that mechanism. They observe a strong dominance of the region 1 FAC generated magnetic fields both at high and low latitudes and diminishing of the Chapman-Ferraro current system. Two sets of spacecraft observations at high and low latitudes suggest that the dayside magnetopause assumes a very blunt shape consistent with the predictions of the MHD simulations. However, *Nakano and Iyemori* [2003] show that during magnetic storms (when all the magnetospheric currents have been developed), the magnitude of region 2 Birkeland currents is comparable with the region 1 currents in postnoon and premidnight sectors and, thus, the contribution of the region 1 FAC to the dayside geomagnetic field is reduced by the region 2 currents. This finding supports the experimental result by *Nagai* [1982], who found that no appreciable variations in the dayside geosynchronous magnetic field are associated with the Birkeland currents during geomagnetic storms.

Modeling the dayside magnetopause by a method of Artificial Neural Network, *Dmitriev and Suvorova* [2000] found a "dimple" arising in the subsolar region when IMF $Bz$ < -10 nT. Following to *Rufenach et al.* [1989] they interpreted the feature as a result of the magnetopause erosion. The dimple is a large-scale local structure. It is unlikely that such structure is formed by the cross-tail and/or field-aligned currents, which produce a global magnetic effect in the dayside hemisphere. Hence, it should be another phenomenon responsible for that distortion.

Using high-resolution 3-D MHD simulations *Borovsky et al.* [2008] found that the reconnection can be saturated by a ''plasmaspheric effect'': high-density magnetospheric plasma flows from plasmasphere into the magnetopause reconnection site and mass loads the reconnection such the reconnection rate is locally reduced in the subsolar region. Based on statistical analysis of the magnetopause crossings by geosynchronous satellites, *Suvorova et al.* [2003] proposed that the reconnection saturation is caused by an enhanced thermal pressure of the magnetospheric plasma and ring current particles during strong magnetic storms. This idea is supported by a recent model of geosynchronous magnetopause crossings (GMCs) [*Dmitriev et al.*, 2011]. Namely, the minimal solar wind pressure required for GMCs under saturation decreases with the storm-time $Dst$-variation, which is produced by the intense ring current.

From hybrid MHD and kinetic multi-fluid simulations, *Winglee et al.* [2008] found that for southward IMF, the subsolar magnetopause is not smooth, as predicted by *Dungey* [1963] or global MHD, but instead it is the rippled blunted surface suggested by the multi-X line models, and this rippling has implications for spacecraft observations. It means that a slow moving spacecraft will have multiple magnetopause crossings even under steady solar wind conditions, with the depth of penetration dependent on the size of the flux rope and its displacement relative to the spacecraft.

Therefore, numerical and empirical models predict that the reconnection saturation under a large southward IMF should be accompanied by prominent distortions of the subsolar magnetopause such as a dimple or rippling. In the present paper, we study the shape of the dayside low-latitude magnetopause with using GMCs occurred during strong magnetospheric disturbances. The method of GMC analysis is described in Section 2. The magnetopause shape during strong southward IMF is analyzed in Section 3. The results are discussed in Section 4. Section 5 gives conclusions.

## 2. Geosynchronous magnetopause crossings

GMCs were collected by using magnetic field data acquired from GOES 8, 9, and 10 geosynchronous satellites in the time interval from 1994 to 2001. The method of GMC identification is described in detail by *Suvorova et al.* [2005]. Briefly, we used high-resolution (~1-min) ISTP data (http://cdaweb.gsfc.nasa.gov/cdaweb/istp_public/) from GOES satellites and solar wind data from upstream monitors Geotail, Wind, and ACE. A magnetopause crossing was identified, when one of two requirements were satisfied: (1) the GOES magnetic field deviated significantly from the geomagnetic field and (2) the $By$ and $Bz$ components measured by the GOES correlated with the corresponding IMF components measured by an upstream monitor.

Figure 1 shows an example of GMC identification by GOES-10 during time interval from 20 to 21 UT on 8 April 2001. In the dayside magnetosphere, the geomagnetic field is characterized by a dominant northward component $Hp$ > 100 nT. When GOES-10 is located in the magnetosheath from 2012 to 2029 UT, from 2043 to 2052 UT, and from 2055 to 2058 UT, the component $H$p turns to southward and the total magnetic field $H$ decreases. The $B$y and $B$z components, measured by GOES-10 in the magnetosheath, demonstrate good correlation with the IMF $B$y and $B$z components measured by ACE upstream monitor with a time delay of 40 min.

The time delay for solar wind propagation was determined independently from cross-correlation between the solar wind total pressure $P$sw and a pressure $P_{\text{Dst}}$, deduced from 1-min SYM-H index (equivalent to $Dst$ variation) with expression $\Delta Dst = b\sqrt{\Delta P_{Dst}}$ [e.g. *Burton et al.*, 1975]. The coefficient $b$ was calculated from a linear regression of the observed $Dst$ and $P$sw. For the present time interval, the cross-correlation between $P$sw and $P_{\text{Dst}}$ was ~0.56 that proved our choice of the time delay. The $P$sw was calculated as the sum of the solar wind dynamic, thermal and magnetic pressures. Note that during strong geomagnetic disturbances the contribution of the thermal and magnetic pressures to the $P$sw can reach up to 30%.

The magnetopause has a systematic distortion related to the Earth rotation around the Sun with orbital speed of ~30 km/s. In order to eliminate the aberration effect, all the coordinates and vector parameters were converted in the aberrated GSM (aGSM) coordinate system. Here we take into account both the orbital velocity of the Earth and non-radial solar wind propagation. The aberration is represented by angles a$Y$ and a$Z$ of rotation about the Z-GSE and aberrated Y-GSE axes, respectively [e.g. *Dmitriev et al.*, 2003].

The orientation of the magnetopause was determined from the rotation of magnetic field across the magnetopause using minimal variance analysis (MVA). Note that the results of MVA vary by the length of the time interval chosen for the analysis. If the results vary greatly for different time intervals, they are unreliable. By varying the boundaries of the time intervals for each GMC, we found time ranges for which the results of MVA changed slightly and gradually (i.e. the solution of MVA was stable). By this way, we calculated a normal to the magnetopause, i.e. local orientation of the magnetopause. In Figure 1, the magnetopause normals have been determined for GMCs at 2013, 2029, and 2043 UT. The GMCs at 2053, 2056 and 2058 UT were very short and, hence, the determination of normals was difficult.

As an example, we consider the normal to the subsolar magnetopause (lat = -7.1°, lon = 1.6°) determined in the time range from 2026 to 2032. The normal was **n** = (0.73, -0.67, 0.03), i.e. it was pointed sunward and dawnward. Using a model by *Lin et al.* [2010] (hereafter reference model) under the current upstream solar wind conditions, we calculated a reference normal $\mathbf{n_r}$ = (0.99, 0.02, -0.09). As expected, the reference normal is mainly directed sunward. Note that the model predicts GMCs very well [*Dmitriev et al.*, 2011]. In addition, the model predicts a north-south asymmetry of the magnetopause related to the tilt angle of geomagnetic dipole. Hence for the reference model, the subsolar point (a point with coordinates aGSM $y = 0$ and $z = 0$ Re) can be different from the nose point, where the normal $\mathbf{n_r}$ = (1., 0, 0).

We can see that the measured normal to the magnetopause is substantially different from the reference one. The unusual dawnward deviation of the normal in the XY plane cannot be explained by the effect of geomagnetic tilt appearing in the XZ plane. The deviation can also result from tilted interplanetary fronts or from time-dependent reconnection and multiple flux transfer events (FTEs) formed during strong southward IMF [*Omidi et al.*, 2009]. For the GMC at 2029 UT, the upstream solar wind did not exhibit any front of interplanetary discontinuity. Hence, the unusual magnetopause orientation might result from FTE occurred during prolonged interval of strong negative $Bz$ of ~ -50 to -100 nT in the magnetosheath as measured by GOES-10. Here we are unable to verify and analyze the FTE effect, because GOES satellites do not provide any plasma data. Instead, we perform a statistical study of deviations of the measured magnetopause normals from the nominal ones.

Figure 2 shows a scatter plot of the collected GMCs in aberrated GSM coordinates. The crossings occupy a wide longitudinal sector from -30° to 50° and latitudinal sector from -25° to 20°. The wide latitudinal spread is due to a superposition of 23.4° tilt of the Earth's rotation axis to the ecliptic plane and ~11° tilt of the geomagnetic dipole relative to the rotation axis. The wide angular spread of GMCs makes possible statistical studies of the shape of dayside low-latitude magnetopause.

### 3. Magnetopause shape at low latitudes

Figure 3 shows projections to the equatorial plane of the magnetopause normals observed during GMCs relative to aGSM X-axis and relative to the reference magnetopause calculated for the solar wind consitions accompanying the GMCs. The orientation is characterized by a wide random scatter. Significant inclinations of the magnetopause normal are revealed in ~15° vicinity of the nose point (mlon = 0 and mlat = 0). As we mentioned above, very strong magnetopause distortions can result from tilted interplanetary fronts and from multiple FTEs formed during strong southward IMF. These effects are beyond the scope of the present study. Hence for convenience, we exclude from further consideration the normals deviated strongly (with a tilt >45°) from the reference normals. As one can see in Figure 3a, the number of GMCs with strongly deviated normals is small and practically all of them are indicated by blue bars, i.e. they occur either under weak southward IMF (Bz > -5 nT) or under very strong solar wind pressures ($P$sw > 21 nPa).

Solar wind and geomagnetic conditions observed by upstream monitors during GMCs are presented in Figure 4. One can see that the $Dst$ variation (see Figure 4a) is mostly negative and large (-310 < Dst <30 nT), i.e. vast majority of GMCs occurs during magnetic storms. The upstream solar wind conditions (see Figure 4b) vary in a very wide dynamic range: -40 < $Bz$ < 40 nT and $P$sw >5 nPa. The conditions are restricted rather sharply by a lower envelope boundary, below which GMCs are not observed. The envelope boundary corresponds to minimal solar wind conditions required for GMCs and can be presented numerically by the following expression [*Suvorova et al.*, 2005]:

$$Psw = 21 - \frac{16.2}{1 + \exp\{0.2(Bz - 2.)\}} \quad (1)$$

The right branch ($Bz \to \infty$), asymptotically approaching $Psw$ = 21 nPa, corresponds to a regime of pressure balance, under which the dayside magnetopause is only driven by solar wind pressure. The left branch ($Bz \to -\infty$) approaches $Psw$ ~4.8 nPa under strong southward IMF and is attributed to the regime of reconnection saturation.

In order to study the magnetopause shape under saturation, we select GMCs accompanied by upstream solar wind conditions with IMF $Bz$ < -5 nT and $Psw$ < 21 nPa. For $Psw$ > 21 nPa, both southward IMF and strong solar wind pressure can cause a GMC and, hence, it is hard to distinguish between the magnetopause effects produced by two different upstream drivers. The threshold of IMF $Bz$ < -5 nT is chosen because of two circumstances. On the one hand, this allows collecting enough statistics in close vicinity and inside the saturation regime. On the other hand, in Figure 1 we demonstrate that during IMF $Bz$ ~ -5 nT and strong solar wind pressure $Psw$ ~ 18 nPa, the southward component of magnetosheath magnetic field, which affects directly the magnetopause, can be extremely large (up to -100 nT) that is considered as a sufficient condition for the reconnection saturation [e.g. *Siscoe et al.*, 2004; *Borovsky et al.*, 2008]. Hence, as a first approach we can attribute the regime of reconnection saturation at geosynchronous orbit to upstream solar wind conditions with IMF $Bz$ < -5 nT and $Psw$ < 21 nPa.

Under saturation, the orientations of normal to the magnetopause exhibit a specific pattern. In the equatorial (X-Y) plane, the normals are mainly tilted dawnward relative to the reference ones in both prenoon and postnoon sectors as one can see in the left panel of Figure 3. Comparing the orientation of normals with aGSM X-axis (right panel of Figure 3), we find that at longitudes from ~5° to 35°, the average orientation is almost parallel to the X-axis. In the prenoon sector, the normals are mainly tilted dawnward. The duskward tilt is found in the late postnoon sector (lon > 40°). In Figure 4a one can see that most of these GMCs occur under not very high solar wind pressures $Psw$ < 15 nPa and during strong magnetic storms, namely at the main phase when $Dst$ < -50 nT. Only three GMCs occur during storm commencement ($Dst$ > 0 nT) related to strong $Psw$.

Statistically, the observed orientation of the normals can be represented as a bluntness extended to the postnoon sector with simultaneous duskward skewing of both prenoon and postnoon wings of the magnetopause. A sketch of the magnetopause cross-section in the aGSM equatorial plane is shown in Figure 5. One can see that under saturation, the equatorial magetopause is characterized by a blunt nose region, which is located in the postnoon sector. In the first approach, the dayside magnetopause might be rather symmetric around a new $X'$-axis, which is shifted toward dusk. Such representation is in a good agreement with properties of the magnetopause dawn-dusk asymmetry reported by *Dmitriev et al.* [2004; 2005].

An obtuse shape of the low-latitude dayside magnetopause under saturation is also revealed in the meridional (X-Z) plane shown in Figure 6. Here we consider separately the prenoon and postnoon sectors. As one can see in the left panel of Figure 6, in the both sectors, the normals in the northern (southern) hemisphere tend to tilt southward (northward) relative to the reference normals that indicate to a larger curvature of the magnetopause than that predicted by the reference model.

An asymmetry between the prenoon and postnoon sector is revealed in the orientations of the magnetopause normal relative to aGSM X-axis in meridional plane, as shown in right panel of Figure 6. In the prenoon sector, the normal demonstrates a regular increase of the tilt with the magnitude of latitude. In the postnoon sector, the orientation of normal exhibits more complicated behavior. Note that for the present case, the statistics of normals is slightly asymmetric relative to the equator: the Southern hemisphere is more represented by GMCs. It seems in the totally aberrated coordinate system, the asymmetry has rather statistical nature than any physical meaning. Moreover in average, the orientation of the normals can be considered to be symmetrical relative to the equator. At latitudes >10°, the normal is predominantly tilted outward from the equator, i.e. poleward that is similar to that observed in the prenoon sector. In the latitudinal ranges from 5° to 10°, the average tilt is close to zero that indicates to the planar magnetopause perpendicular to the X-axis.

At equatorial latitudes <5°, we find that 4 out of 5 normals are tilted toward the equator that indicates to a negative curvature of the magnetopause, or dimple. Table 1 lists the basic characteristics of the GMCs in the dimple region. The GMCs occurred during 3 different magnetic storms, under very strong southward IMF $Bz$ < -14 nT and at various aGSM longitudes from ~6° to 43°. GMCs #1 and 3 were accompanied by extremely strong southward IMF ($Bz$ < -20 nT), which is definitely proper for the reconnection saturation. For the "abnormal" GMC #5, whose normal is tilted outward from the equator, Table 1 shows the weakest magnitude of negative $Bz$ = -14.3 nT and very large solar wind pressure $Psw$ = 19.6 nPa. It might be that for the present case, the effect of very strong pressure masked the effect of relatively weak southward IMF.

Figure 7 shows a sketch of meridional cut of the magnetopause in the postnoon sector. In contrast to the reference model, the shape of magnetopause is not smooth. In general, it is characterized by a larger curvature. But at low latitudes, the curvature increases infinitely (plane) and then takes a negative value that implies a dimple at equatorial latitudes. The dimple-like orientation of the normals appears in a wide range of longitudes. Geometrically, this feature can be described as an equatorial trench at the postnoon magnetopause.

**Discussion**

From GOES observations of the magnetopause crossings during reconnection saturation, we have found prominent distortions of the magnetopause in comparison with the reference model by *Lin et al.* [2010]. Figures 5 and 7 show the sketches of magnetopause cross-section in the equatorial and meridional planes, respectively. We can combine the orthogonal cross-sections into a 3-dimensional sketch of the dayside magnetopause as shown in Figure 8. An important feature of the resulting 3-D picture is equatorial trench at the magnetopause in the postnoon sector. In the prenoon sector, the magnetopause is blunted and smooth. At low latitudes, the magnetopause is shifted duskward such that the nose region occurs in the postnoon sector.

Note that the sketches presented in Figures 5 and 7 are based on limited statistics. Here we have to remind that geosynchronous crossings of the magnetopause at low latitudes in vicinity of noon are very rare because of two circumstances [*Suvorova et al.*, 2005]. In the noon region, geosynchronous satellites spend most of time at middle latitudes (>20°) and, in addition, the solar wind conditions required for GMCs are quite infrequent. Hence, in order to collect "sufficient" amount of observations, which can definitely prove the results, we need to accumulate several decades of continuous data. Up to date, only ~10 years of continuous observations of GMCs and solar wind conditions are available. Based on these limited data, we found a pattern of magnetopause features, which were observed independently during different time intervals and at different longitudes. The wide spatial and temporal ranges support the validity of limited statistics. In addition, the revealed pattern does not contradict to the existing findings and complements the previous expectations.

It is commonly accepted that the magnetopause, a current layer separating the magnetospheric magnetic field from the interplanetary medium, is formed in interaction of the solar wind with the geomagnetic dipole. The interaction results in generation of a complex geomagnetic field and a system of geomagnetic currents. Hence, the solar wind plasma streams and IMF are primary external drivers of the magnetopause and geomagnetic field [e.g. *Parker*, 1996; *Vasyliunas*, 2001]. Under reconnection regime, the magnetosphere states in a dynamic equilibrium when the effect of strong southward IMF as external driver of the magnetopause is counteracted by effects of the magnetospheric origin, which we can call internal magnetospheric drivers.

Discussing possible drivers of the magnetopause dawn-dusk asymmetry, *Dmitriev et al.* [2004] show that asymmetries in the upstream solar wind drivers are unlikely responsible for the duskward skewing of the magnetopause. It was concluded that the storm-time partial ring current is rather a source of the dawn-dusk asymmetry. The thermal pressure of hot plasma populating the storm-time ring current was also proposed as a force, which restricts and saturates the magnetic reconnection at the magnetopause [*e.g. Dmitriev et al.*, 2011]. However, the effects of ring current and thermal pressure can hardly explain the appearance of magnetopause bluntness and trench.

*Siscoe et al.* [2004] proposed that the region 1 FACs result in distortions at the dayside magnetopause. However, during magnetic storms the region 2 FACs can be intense such that the net contribution of the Birkeland currents to the dayside magnetic field is small [*Nakano and Iyemori*, 2003; *Nagai*, 1982]. In addition to the Birkeland and ring currents, the cross-tail current contributes to the geomagnetic field. During magnetic storms, the cross-tail current is very intense and its magnetic effect at the dayside magnetopause might become comparable with the geodipole field [*Maltsev et al.*, 1996; *Alexeev et al.*, 2001; *Turner et al.*, 2000].

The depletion of the dayside geomagnetic field by the cross-tail current is well established [*Maltsev et al.*, 1996; *Alexeev et al.*, 1996]. This depletion results from storm-time intensification of the current and from sunward motion of its inner edge such that the enhanced current approaches to the dayside magnetopause. The geometry of magnetic field, produced by the storm-time cross-tail current at the magnetopause, is apparently different from the dipole. The field is rather close to a linear configuration formed near the edge of a large-scale flat electric current. The flat current produces almost constant magnetic effect at different latitudes. The negative magnetic effect of the "linear" magnetic field should be biggest near the magnetic equator, where the dipole magnetic field is weakest. Hence, it is reasonably to propose that the bluntness of the dayside magnetopause is related to a strong contribution of the cross-tail current. But the trench at the equatorial magnetopause is hardly related to the global magnetic effect of the tail current.

The equatorial trench in the nose region might be produced by the strong region 1 FAC, when the region 2 FAC is still undeveloped, i.e. in the beginning of magnetic storm. Such conditions might appear for GMC #4 from Table 1. However for most GMCs, the magnetopause is characterized by a prominent dawn-dusk asymmetry that testifies to the strong partial ring current and a well-developed magnetic storm activity on the main phase. Hence, the storm-time formation of the trench can be explained hardly by the effect of region 1 FAC.

Under storm conditions, the equatorial trench might result from a large-scale quasi-state reconnection at the dayside magnetopause. During the reconnection, the magnetic flux tubes are transferred from the equatorial magnetopause to the tail. This transport produces an indentation at the low-latitude magnetopause and an expansion of the magnetopause at middle and high latitudes. For very large negative $B_z$, the reconnection is mostly effective in the nose region. We have shown that under saturation, the nose region is shifted in the postnoon sector. Hence, the strong reconnection at the dayside magnetopause might be a possible mechanism of the trench formation in the postnoon sector. Under saturation, the equatorial magnetopause deflation, produced by reconnection, can be balanced by a thermal pressure of the hot plasma from the strong ring current. Further experimental studies of the plasma and particles near the magnetopause during strong magnetic storms are required in order to verify this scenario.

**Conclusions**

GOES observations of the magnetopause at geosynchronous orbit allow determining the shape of dayside magnetopause under saturation related to the effect of strong negative $Bz$. We found a prominent dawn-dusk asymmetry consisting in duskward shifting of the magnetopause such that the nose region is located in postnoon sector. At low latitudes, the magnetopause is blunt. In the postnoon sector, a trench is formed at the equatorial magnetopause.

The effects of reconnection saturation and duskward skewing can be explained by a substantial contribution of the hot magnetospheric plasma from the strong asymmetrical ring current into the pressure balance at the dayside magnetopause during magnetic storms. The bluntness can result from the magnetic effect of the intensified cross-tail current. The origin of equatorial trench in the postnoon region might be related to a large-scale quasi-state reconnection at the dayside magnetopause.

**Acknowledgements**

This work was supported by grants NSC-99-2111-M-008-013 and NSC-100-2111-M-008-016 from the National Science Council of Taiwan and by Ministry of Education under the Aim for Top University program at National Central University of Taiwan.

Table 1. Characteristics of GMCs in the trench region

| # | Date and UT time | mlat, mlon, deg | $PS$, deg | $Dst$, nT | $Bz$, nT | $P$sw, nPa |
|---|---|---|---|---|---|---|
| 1 | 2000/4/06 1835 | -4.7, 34.4 | 17 | -94 | -23.37 | 9.13 |
| 2 | 2001/3/31 1822 | 3.5, 32.8 | 14 | -268 | -17.53 | 15.36 |
| 3 | 2001/3/31 1651 | -1.3, 6.6 | 15 | -243 | -28.65 | 5.14 |
| 4 | 2001/8/17 1937 | -1.7, 42.8 | 21 | -47 | -18.67 | 20.37 |
| 5 | 2001/8/17 1947 | -2.1, 42.0 | 21 | -55 | -14.25 | 19.57 |

**Figure Captions**

Figure 1. An example of magnetopause crossing identification by GOES-10 magnetic data from 20 to 21 UT on 8 April 2001 (from top to bottom): GOES-10 horizontal $H$p (black solid curve) and total $H$ (blue dotted curve) magnetic field; solar wind pressure $P$sw (black solid curve) and *Dst*-deduced pressure $P_{Dst}$ (blue dotted curve); aberrated GSM (aGSM) $B$z, $B$y, and $B$x components of magnetic filed measured by ACE (black solid curves) and GOES-10 (blue dotted curves); solar wind aberration angles a$Y$ (black solid curve) and a$Z$ (blue dotted curve), aGSM latitude mLat and local time MLT. The $B$z and $B$y components measured by GOES are divided by 10 and $B$x component is divided by 5. The vertical red dotted lines indicate magnetopause crossing. The vertical blue dashed-dotted lines restrict the time interval for MVA analysis.

Figure 2. Location of geosynchronous magnetopause crossings (GMCs) observed by GOES in aGSM coordinates. The red dots correspond to IMF $B$z < -5 nT and $P$sw <21 nPa.

Figure 3. Projection of the magnetopause normal, observed during GMCs, to the equatorial plane for all events (blue and red segments) and for those during the reconnection saturation for IMF $B$z < -5 nT and $P$sw < 21 nPa (red segments). Left and right panels show the orientation, respectively, relative to the reference magnetopause normal **n**<sub>r</sub> and relative to aGSM X-axis. The magnetopause is characterized by a strong dawn-dusk asymmetry and a prominent bluntness in the postnoon sector.

Figure 4. Scatter plots of the solar wind pressure $P$sw during GMCs versus a) *Dst* variation, and b). IMF $B$z in aGSM coordinates The horizontal dashed line indicates the $P$sw = 21 nPa. The thick solid curve indicates an envelope boundary of necessary conditions for GMCs [*Suvorova et al.*, 2005]. The red circles correspond to the conditions proper for reconnection saturation.

Figure 5. A sketch of the magnetopause cross-section in the aGSM equatorial plane: for the reference magnetopause (dashed black curve) and the magnetopause under saturation (solid red curve). For the latter case, the magnetopause shape is characterized by a prominent bluntness in the nose region shifted to the postnoon sector. The new axis of symmetry X' of the magnetopause is denoted by the red solid arrow, which is shifted duskward from the nominal *X*-axis denoted by the black dashed arrow.

Figure 6. Projections of the magnetopause normal to the meridional plane observed under reconnection saturation ($B$z < -5 nT and $P$sw < 21 nPa): (left) in comparing with the normals to the reference magnetopause and (right) relative to the aGSM *X*-axis in the prenoon (blue segments) and postnoon sectors (red segments). In the left panel, the orientation of normals exhibits a strong bluntness of the magnetopause. In the right panel, the pattern of deviations in the postnoon sector indicates a dimple at low-latitudes.

Figure 7. A sketch of the magnetopause cross-section in the aGSM meridional plane in the postnoon sector: for the reference magnetopause (dashed black curve) and the magnetopause under saturation (solid red curve). For the latter case, a dimple is formed at the low-latitude magnetopause.

Figure 8. A sketch of the dayside magnetopause under saturation. The magnetopause is skewed duskward. In the afternoon sector, a trench is formed at low latitudes.

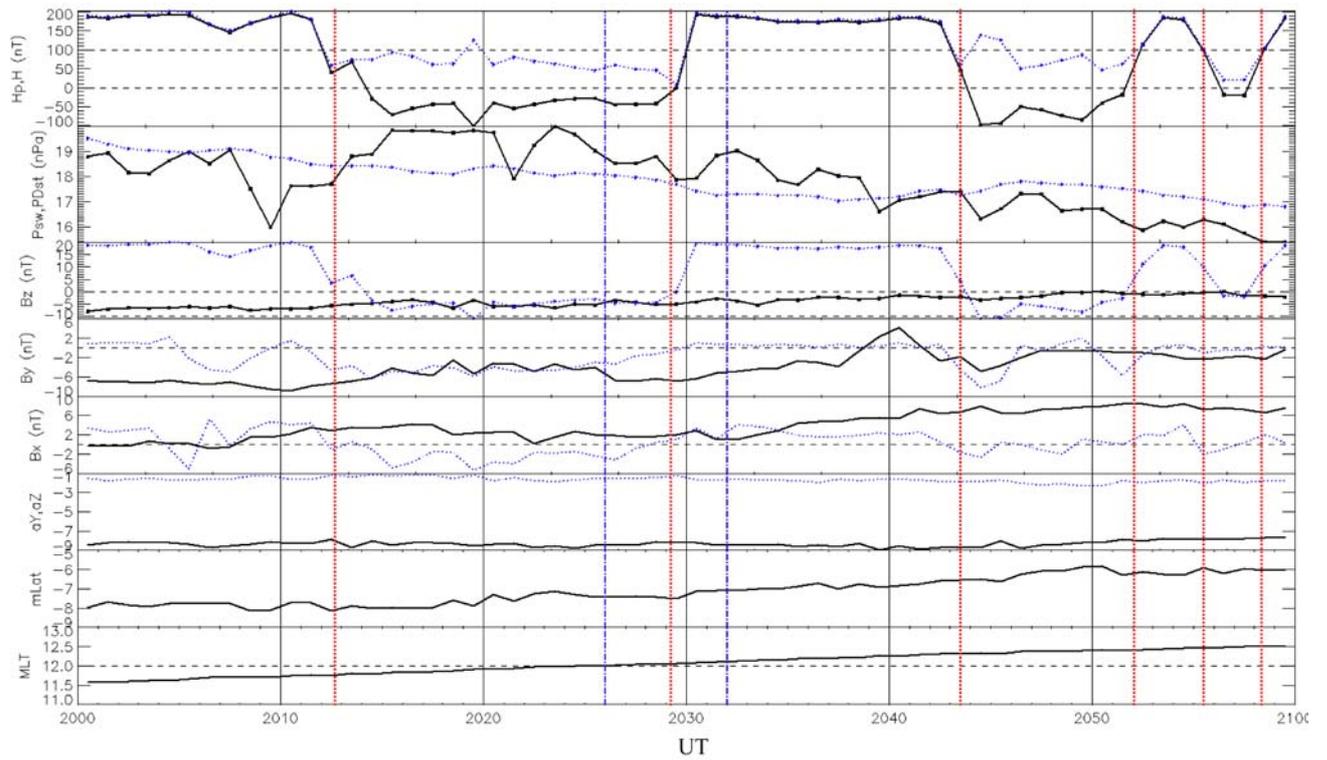

Figure 1. An example of magnetopause crossing identification by GOES-10 magnetic data from 20 to 21 UT on 8 April 2001 (from top to bottom): GOES-10 horizontal $H$p (black solid curve) and total $H$ (blue dotted curve) magnetic field; solar wind pressure $P$sw (black solid curve) and $Dst$-deduced pressure $P_{Dst}$ (blue dotted curve); aberrated GSM (aGSM) $B$z, $B$y, and $B$x components of magnetic filed measured by ACE (black solid curves) and GOES-10 (blue dotted curves); solar wind aberration angles a$Y$ (black solid curve) and a$Z$ (blue dotted curve), aGSM latitude mLat and local time MLT. The $B$z and $B$y components measured by GOES are divided by 10 and $B$x component is divided by 5. The vertical red dotted lines indicate magnetopause crossing. The vertical blue dashed-dotted lines restrict the time interval for MVA analysis.

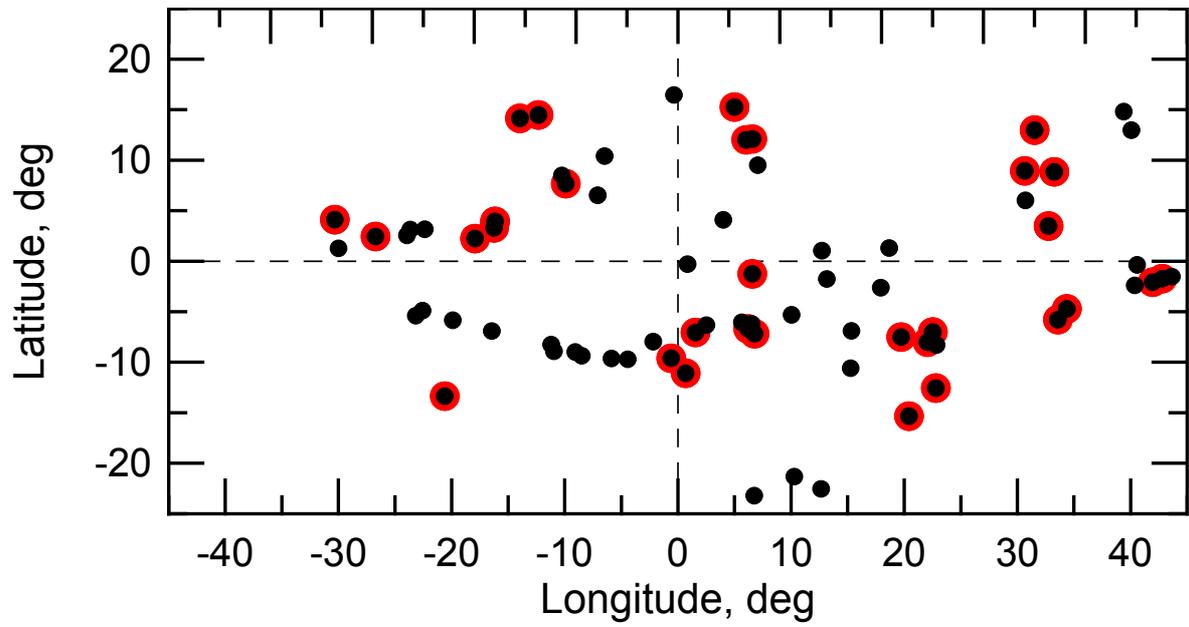

Figure 2. Location of geosynchronous magnetopause crossings (GMCs) observed by GOES in aGSM coordinates. The red dots correspond to IMF $B$z < -5 nT and $P$sw <21 nPa.

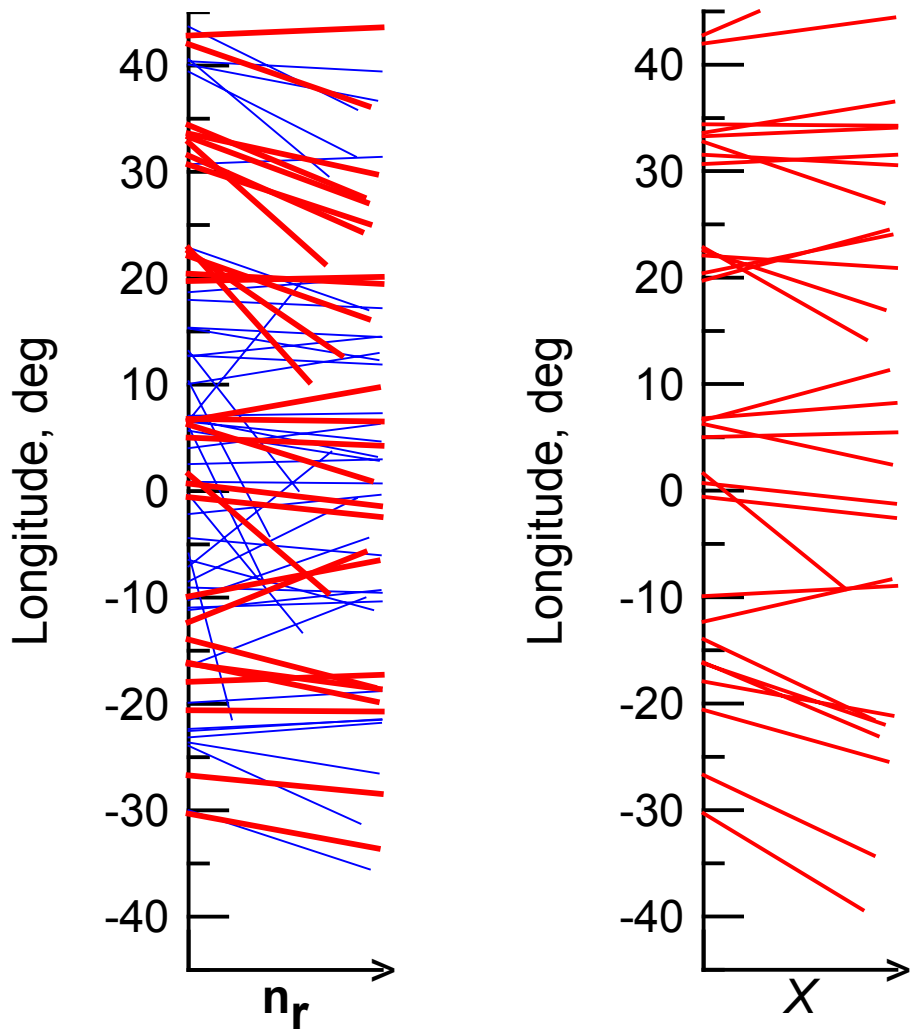

Figure 3. Projection of the magnetopause normal, observed during GMCs, to the equatorial plane for all events (blue and red segments) and for those during the reconnection saturation for IMF $Bz < -5$ nT and $P$sw $< 21$ nPa (red segments). Left and right panels show the orientation, respectively, relative to the reference magnetopause normal $\mathbf{n_r}$ and relative to aGSM X-axis. The magnetopause is characterized by a strong dawn-dusk asymmetry and a prominent bluntness in the postnoon sector.

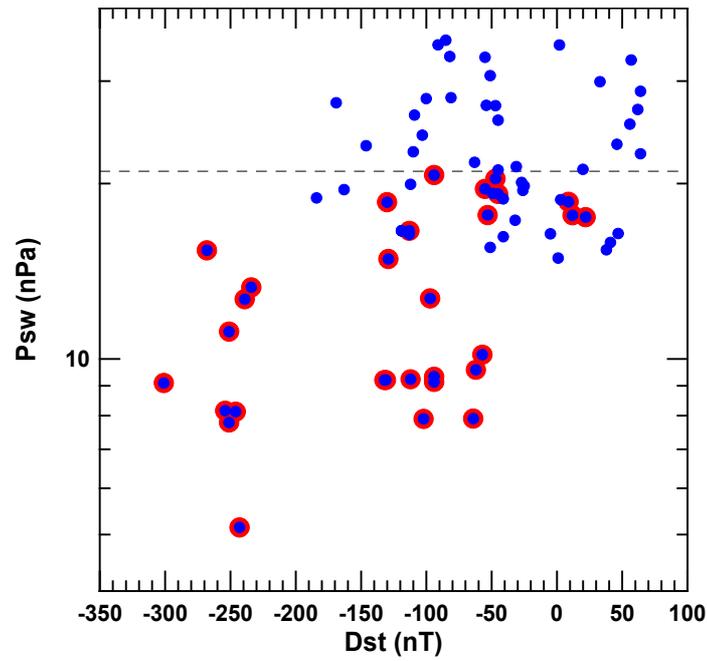

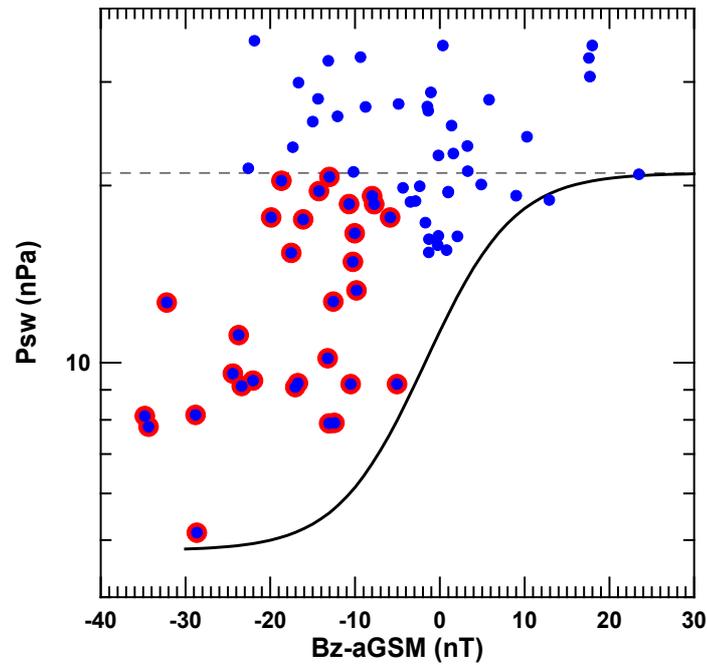

Figure 4. Scatter plots of the solar wind pressure Psw during GMCs versus a) *Dst* variation, and b). IMF *B*z in aGSM coordinates The horizontal dashed line indicates the *P*sw = 21 nPa. The thick solid curve indicates an envelope boundary of necessary conditions for GMCs [*Suvorova et al.*, 2005]. The red circles correspond to the conditions proper for reconnection saturation.

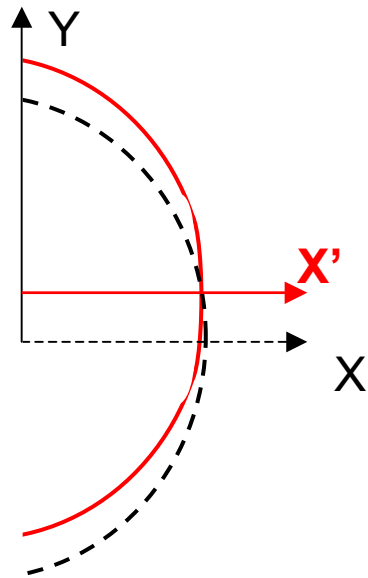

Figure 5. A sketch of the magnetopause cross-section in the aGSM equatorial plane: for the reference magnetopause (dashed black curve) and the magnetopause under saturation (solid red curve). For the latter case, the magnetopause shape is characterized by a prominent bluntness in the nose region shifted to the postnoon sector. The new axis of symmetry *X'* of the magnetopause is denoted by the red solid arrow, which is shifted duskward from the nominal *X*-axis denoted by the black dashed arrow.

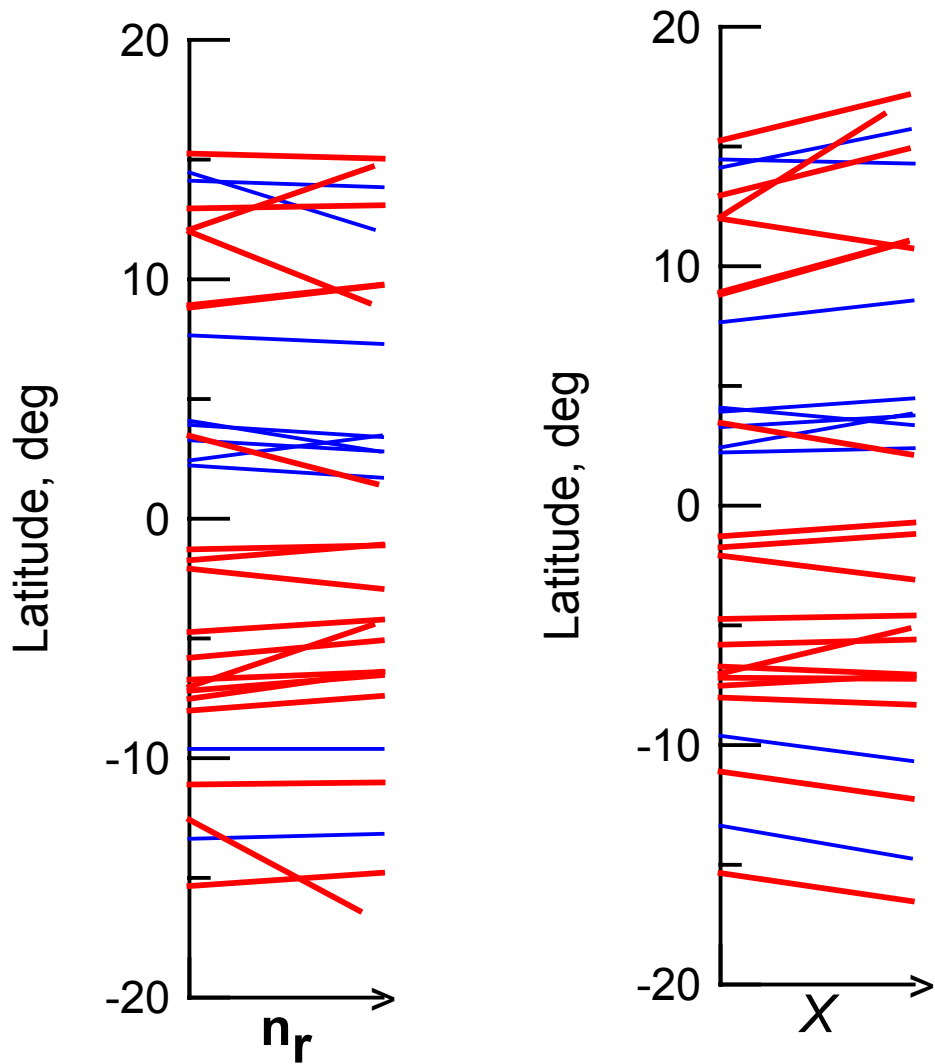

Figure 6. Projections of the magnetopause normal to the meridional plane observed under reconnection saturation ($Bz < -5$ nT and $P$sw $< 21$ nPa): (left) in comparing with the normals to the reference magnetopause and (right) relative to the aGSM $X$-axis in the prenoon (blue segments) and postnoon sectors (red segments). In the left panel, the orientation of normals exhibits a strong bluntness of the magnetopause. In the right panel, the pattern of deviations in the postnoon sector indicates a dimple at low-latitudes.

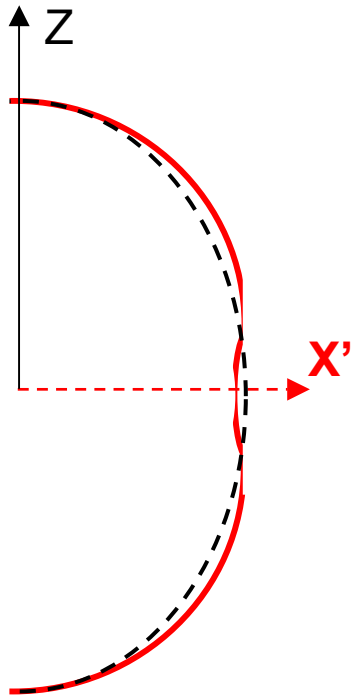

Figure 7. A sketch of the magnetopause cross-section in the aGSM meridional plane in the postnoon sector: for the reference magnetopause (dashed black curve) and the magnetopause under saturation (solid red curve). For the latter case, a dimple is formed at the low-latitude magnetopause.

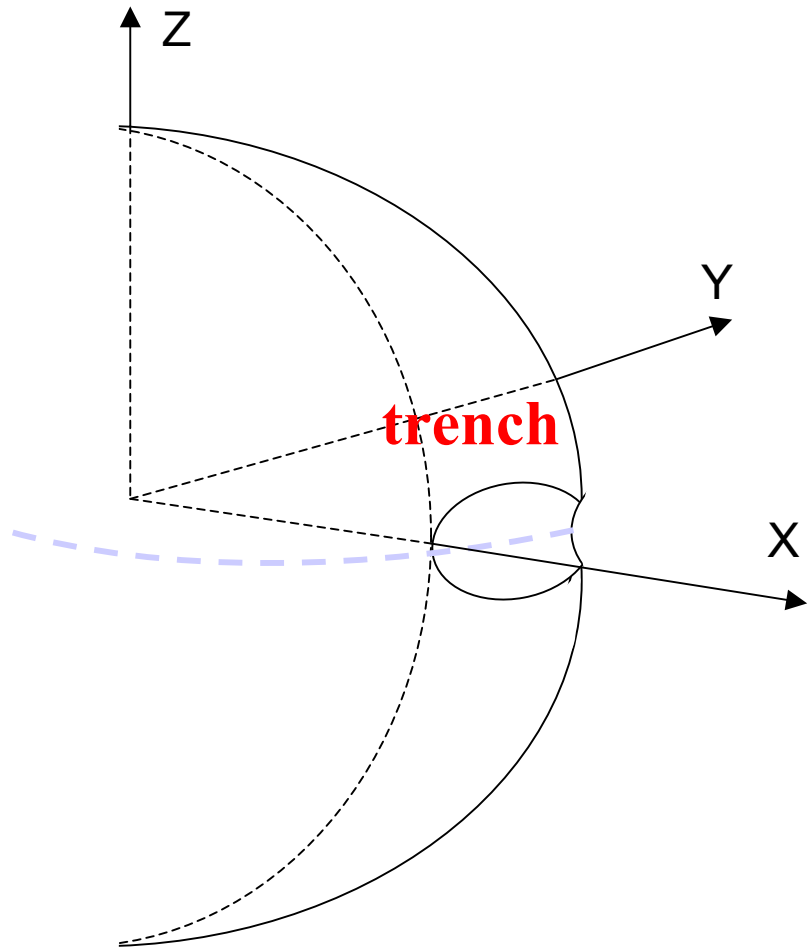

Figure 8. A sketch of the dayside magnetopause under saturation. The magnetopause is skewed duskward. In the afternoon sector, a trench is formed at low latitudes.